\documentclass[preprint,showpacs,showkeys,preprintnumbers,amsmath,amssymb,aip,apl]{revtex4-1}

\usepackage{graphicx}
\usepackage{float}
\usepackage{color}

\begin{document}

\preprint{Kozina et al, NiTiSn}

\title{Electronic structure and symmetry of valence states of epitaxial
NiTiSn and NiZr$_{0.5}$Hf$_{0.5}$Sn thin films by hard x-ray photoelectron spectroscopy.}

\author{Xeniya Kozina}
\affiliation{Institut f{\"u}r Anorganische und Analytische Chemie,
             Johannes Gutenberg - Universit{\"a}t, 55099 Mainz, Germany.}
\author{Tino Jaeger}
\affiliation{Institut f{\"u}r Physik,
             Johannes Gutenberg - Universit{\"a}t, 55099 Mainz, Germany.}
\author{Siham Ouardi}
\affiliation{Institut f{\"u}r Anorganische und Analytische Chemie,
             Johannes Gutenberg - Universit{\"a}t, 55099 Mainz, Germany.}
\author{Andrei Gloskowskij}
\affiliation{Institut f{\"u}r Anorganische und Analytische Chemie,
             Johannes Gutenberg - Universit{\"a}t, 55099 Mainz, Germany.}
\author{Gregory Stryganyuk}
\affiliation{Institut f{\"u}r Anorganische und Analytische Chemie,
             Johannes Gutenberg - Universit{\"a}t, 55099 Mainz, Germany.}           
\author{Gerhard Jakob}
\affiliation{Institut f{\"u}r Physik,
             Johannes Gutenberg - Universit{\"a}t, 55099 Mainz, Germany.}
\author{Takeharu Sugiyama}
\affiliation{Japan Synchrotron Radiation Research Institute (JASRI), SPring-8, Hyogo 679-5198, Japan}
\author{Eiji Ikenaga}
\affiliation{Japan Synchrotron Radiation Research Institute (JASRI), SPring-8, Hyogo 679-5198, Japan}
\author{Gerhard~H.~Fecher}
\email{fecher@uni-mainz.de}
\affiliation{Institut f{\"u}r Anorganische und Analytische Chemie,
             Johannes Gutenberg - Universit{\"a}t, 55099 Mainz, Germany.}
\affiliation{Max Planck Institute for Chemical Physics of Solids,
             01187 Dresden, Germany.}
\author{Claudia~Felser}
\affiliation{Institut f{\"u}r Anorganische und Analytische Chemie,
             Johannes Gutenberg - Universit{\"a}t, 55099 Mainz, Germany.}
\affiliation{Max Planck Institute for Chemical Physics of Solids,
             01187 Dresden, Germany.}

\date{\today}

\begin{abstract}

The electronic band structure of thin films and superlattices made of Heusler 
compounds with NiTiSn and NiZr$_{0.5}$Hf$_{0.5}$Sn composition was studied by 
means of polarization dependent hard x-ray photoelectron spectroscopy. The 
linear dichroism allowed to distinguish the symmetry of the valence states of 
the different types of layered structures. The films exhibit a larger amount of 
{\it "in-gap"} states compared to bulk samples. It is shown that the films and 
superlattices grown with NiTiSn as starting layer exhibit an electronic 
structure close to bulk materials.

\end{abstract}



\maketitle

%\section{Introduction} %%%%%%%%%%%%%%%%%%%%%%%%%%%%%%%%%%%%%%%%%

The progressively growing interest in exploration and design of the materials 
exhibiting thermoelectric properties is mediated by their potential applications 
in new environment friendly industrial technologies for power generation and 
refrigeration~\cite{Sootsman2009}. As the efficiency of a thermoelectric device 
solely depends on the dimensionless figure of merit $ZT=S^2\sigma\kappa^{-1}$ at 
operating temperature, the most interesting materials are those with high $ZT$, 
which is, in turn, defined by thermopower $S$, electric conductivity $\sigma$ 
and thermal conductivity $\kappa$. Due to the unique tunability of properties, 
thermal and chemical stability, non-toxicity and ease in synthesis, among other 
half-Heusler compounds the Ni$X$Sn based family of compounds and their solid 
solutions have become the most perspective ones for reaching high $ZT$ 
values~\cite{Aliev1990,Sakurada2005,Shutoh2005,Chaput2006}. Many attempts were 
made towards optimization of $ZT$ via enlarging either $S$ or 
{$\sigma$}~\cite{Shutoh2005,Schwall2011}. Alternetively a reduction of $\kappa$ 
allows significantly to rise $ZT$ values, as it was demonstrated for 
YX$_{0.5}$X'$_{0.5}$Z family of half-Heusler compounds~\cite{Hohl1999,Shen2001,Sakurada2005}.

Boundary scattering of electrons and phonons play a major role in further 
suppression of the thermal conductivity in polycrystalline 
materials~\cite{Savvides1980} and thin film superlattices. In the latter case 
the phonons are scattered at the superlattice interfaces when their mean free 
path is shorter than the period of the superlattice leading to low values of the 
cross-plane $\kappa$~\cite{Yanga2002}. Improvement of the quality of such 
multilayer stacks as it was previously demonstrated for epitaxial 
NiTiSn/NiZr$_{0.5}$Hf$_{0.5}$Sn superlattices~\cite{Jaeger2011} will create new 
options for producing high performance thermoelectric devices.

To improve the transport properties of the materials it is necessary to 
understand and explore their electronic structure close to the Fermi energy 
($\epsilon\rm_F$). Hard x-ray photoelectron spectroscopy (HAXPES) is a powerful 
method to probe both chemical states and electronic structure of bulk materials 
and buried layers in a non-destructive way~\cite{Fecher2008,Kozina2010}. The 
combination of HAXPES with polarized radiation for excitation significantly 
extends its applicability. The use of linearly $s$ and $p$ polarized light in 
HAXPES enables the analysis of the symmetry of bulk electronic 
states~\cite{Ouardi2011}. In the present study the valence band electronic 
structure of NiTiSn/NiZr$_{0.5}$Hf$_{0.5}$Sn superlattices were investigated by 
means of HAXPES and linear dichroism.

%\section{Experimental details} %%%%%%%%%%%%%%%%%%%%%%%%%%%%%%%%%%%%%%%%%

\begin{figure}
\centering
   \includegraphics[width=7cm]{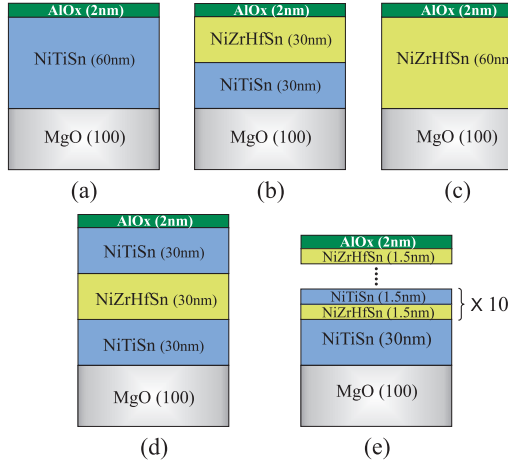}
   \caption{Sketch of the sample structures.
            The layers in (a), (b) and (c) correspond to the 30-nm-thick
            films of NiTiSn and NiZr$_{0.5}$Hf$_{0.5}$Sn compounds grown
            on different buffer layers. (d) presents a bilayer sample and
            (e) shows the  superlattice.}
\label{fig:sample_structure}
\end{figure}

For the present study, multilayer stacks consisting of alternating NiTiSn and 
NiZr$_{0.5}$Hf$_{0.5}$Sn layers were deposited by means of dc-sputtering. The 
details of fabrication and characterization of the samples are described in 
Reference~\cite{Jaeger2011}. Sketches of the investigated thin film, bilayer, 
and superlattice samples are shown in Fig.~\ref{fig:sample_structure}. The 
topmost AlO$\rm_X$ layer serves as a protective cap preventing the oxidation and 
degradation of the thin films.

% Figure 1 %%%%%%%%%%%%%%%%%%%%%%%%%%%%%%%%%%%%%%%%%%%%%%%%%%%%%%%%%
%%%%%%%%%%%%%%%%%%%%%%%%%%%%%%%%%%%%%%%%%%%%%%%%%%%%%%%%%%%%%%%%%%%%

The HAXPES experiment was performed at BL47XU of Spring-8 (Japan) using 
7.940\,keV linearly polarized photons for excitation. Vertical ($s$) direction 
of polarization was achieved by means of a in-vacuum phase retarder based on a 
600-${\mu}$m-thick diamond crystal with a degree of polarization above 90\,\%. 
Horizontal ($p$) polarization was obtained directly from the undulator without 
any additional polarization optics. The energy resolution was set to 250\,meV 
and was verified by spectra of the Au valence band at the $\epsilon\rm_F$. 
Gracing incidence -- normal emission geometry was used ($\theta$=2$^\circ$) that 
ensures that the polarization vector was nearly parallel ($p$) or perpendicular 
($s$) to the surface normal. For further details on HAXPES experiment 
see~\cite{Ouardi2011,Kozina2011}.

%\section{Results and discussion} %%%%%%%%%%%%%%%%%%%%%%%%%%%%%%%%%%%%%%%%%

\begin{figure}
\centering
   \includegraphics[width=5.5cm]{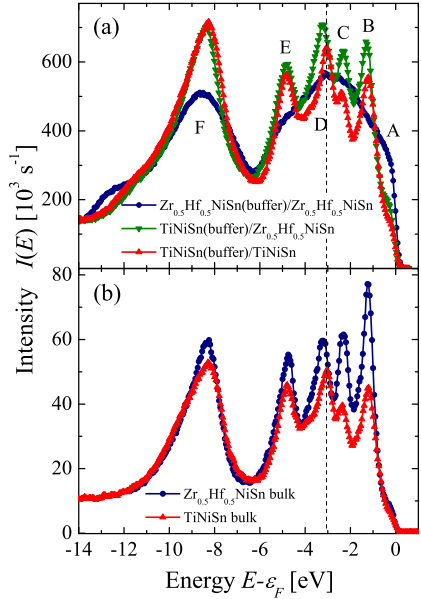}
   \caption{Valence band spectra of the single
            NiTiSn and NiZr$_{0.5}$Hf$_{0.5}$Sn films grown on different buffer
            layers (a) compared to polycrystalline NiTiSn and NiZr$_{0.5}$Hf$_{0.5}$Sn bulk samples (b).
            (Note that the additional intensity at below -10~eV seen in a) emerges from
             the AlO$_x$ cap layer.)}
\label{fig:diff_buff}
\end{figure}

Fig.~\ref{fig:diff_buff}(a) presents the valence band spectra of NiTiSn and 
NiZr$_{0.5}$Hf$_{0.5}$Sn 30-nm-thick films grown on different buffer layers 
(NiTiSn or NiZr$_{0.5}$Hf$_{0.5}$Sn (see Fig.~\ref{fig:sample_structure} (a), 
(b) and (c))). The spectra of the materials grown on a NiTiSn buffer reveal 
clearly narrow structures originating from the band structure of the Heusler 
compounds. The structure at lower binding energies corresponds to the $d$-
states. They are separated by the intrinsic Heusler $sp$-hybridization gap (at 
about -6\,eV) from the $s$-states. In the range above -6\,eV the 4-peak-
structure peculiar to Ni$X$Sn compounds is clearly resolved. Such shape of the 
energy distribution curve is formed mainly by the partial density of Ni-$3d$ 
states as was shown previously (see References~\cite{Ouardi2011,Tobola1998,Pierre1997} 
for the calculated density of states 
(DOS)). The contribution of the Sn $s$ states gives rise to the broad peak at
-8.26\,eV (peak F). Apparently the intensity of $s$-states becomes comparable 
with that of $d$-states at about 8\,keV excitation energy. Such a behavior is a 
direct consequence of different cross sections for $s$ and $d$ states.

% Figure 2 %%%%%%%%%%%%%%%%%%%%%%%%%%%%%%%%%%%%%%%%%%%%%%%%%%%%%%%%%
%%%%%%%%%%%%%%%%%%%%%%%%%%%%%%%%%%%%%%%%%%%%%%%%%%%%%%%%%%%%%%%%%%%%

The peaks positions of NiTiSn and NiZr$_{0.5}$Hf$_{0.5}$Sn films grown on a 
NiTiSn buffer agree well with those of polycrystalline NiTiSn and 
NiZr$_{0.5}$Hf$_{0.5}$Sn bulk samples shown in Fig.~\ref{fig:diff_buff}(b). As 
it was demonstrated before~\cite{Miyamoto2008,Ouardi2011}, the intensity of 
peaks B (-1.21\,eV) and C (-2.41\,eV) undergoes drastic changes. When going from 
NiTiSn to NiZr$_{0.5}$Hf$_{0.5}$Sn, i. e. by substitution of Ti atoms by 
(Zr,Hf), peaks B and C are increased. This follows from the fact that the Ti 
3$d$ partial DOS contributes significantly to the total DOS in this energy range 
along with the Ni 3$d$ states. Larger cross sections for Zr 4$d$ and Hf 5$d$ 
states compared to the Ti 3$d$ states enhance the peaks. Moreover, feature D 
shifts towards higher binding energies by 0.21\,eV in the spectra of both bulk 
samples and thin films when Ti is substituted by (Zr,Hf). This correlation in 
the spectra of the epitaxially grown thin films and the pure polycrystalline 
samples together with the agreement with previously reported results implies the 
formation of a well ordered crystalline C$_{\rm 1b}$ structure in the films of 
both compounds when grown on a NiTiSn buffer layer.

For both compounds one observes the appearance of {\it "in-gap"} states close to 
$\epsilon_{\rm F}$ (feature A). Substitution of Ti atoms with (Zr,Hf) leads to 
an increase of {\it "in-gap"} states in both thin films and bulk samples that is 
in a good agreement with recent work~\cite{Miyamoto2008}. Such states are 
attributed to the disorder at the Ti-site, viz. formation of the antisites of Ti 
atoms with the vacancies~\cite{Ouardi2010}. They are responsible for the 
remarkable thermoelectric properties of the materials. The relatively high 
amount of {\it "in-gap"} states in the thin films compared with the bulk 
materials can be explained by the presence of additional crystalline defects in 
the thin films induced by lattice strain, interface states with broken symmetry, 
or interdiffusion of atoms in conjugated layers. NiZr$_{0.5}$Hf$_{0.5}$Sn grown 
on a NiZr$_{0.5}$Hf$_{0.5}$Sn buffer (Fig.~\ref{fig:diff_buff}(a)) has obviously 
a high degree of disorder as is revealed from both smeared out valence band and 
completely closed band gap.

\begin{figure}
\centering
   \includegraphics[width=5.5cm]{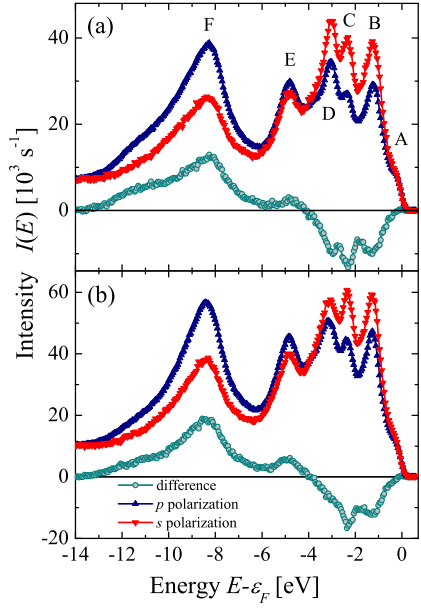}
   \caption{Polarization-dependent valence band spectra of
            a NiZr$_{0.5}$Hf$_{0.5}$Sn/NiTiSn bilayer (a) and the
            NiTiSn/NiZr$_{0.5}$Hf$_{0.5}$Sn superlattice (b). The spectra obtained
            with $s$ and $p$ polarized x-rays are shown together with the difference curves. }
\label{fig:LDAD}
\end{figure}

Further investigations were performed on bilayers and superlattices 
(Fig.~\ref{fig:sample_structure} (d), (e)). Both, $p$- and $s$-polarized, hard 
x-rays were used for excitation. The photoelectron spectra of both samples 
(Fig.~\ref{fig:LDAD}) are typical for the electronic structure of the compounds, 
as described above. The high probing depth in the order of tens of nanometers 
allows to obtain the information from several 1.5-nm-thick layers of the 
superlattice. Their contribution to the total signal is nonequivalent as is seen 
in Fig.~\ref{fig:LDAD}. One notices a relative redistribution of peaks B, C, and 
D when comparing the spectra taken with both orthogonal polarization. A clear 
enhancement of the signals from the B and C states -- similar to the 
NiZr$_{0.5}$Hf$_{0.5}$Sn sample -- is explained by the presence of the topmost 
1.5~nm-thick NiZr$_{0.5}$Hf$_{0.5}$Sn layer in the superlattice. Here, most of 
the obtained signal is attributed to the NiZr$_{0.5}$Hf$_{0.5}$Sn layer whereas 
the intensity from the underlying and other layers is damped due to increased 
inelastic scattering probability for electrons passing larger distances through 
the upper layers of the structure.

% Figure 3 %%%%%%%%%%%%%%%%%%%%%%%%%%%%%%%%%%%%%%%%%%%%%%%%%%%%%%%%%
%%%%%%%%%%%%%%%%%%%%%%%%%%%%%%%%%%%%%%%%%%%%%%%%%%%%%%%%%%%%%%%%%%%%

The spectra shown in Fig.~\ref{fig:LDAD} were normalized to the secondary 
electron background at about -14\,eV to account for different intensities for 
different kind of polarization (see also~\cite{Ouardi2011}). Substantial changes 
of the spectra from both samples are quite obvious when the polarization is 
switched from $p$ to $s$. In both cases the peak at -8.31\,eV arising from Sn 
$s$ ($a1$) states is enhanced with $p$-polarized photons, while the intensity of 
the $d$-part of the spectra is lowered. Namely the features originating from $e$ 
and $t_{2}$ states (-2.36\,eV) and $t_{2}$ states (-3.06\,eV) of Ni as well as 
$e$ and $t_{2}$ states (-1.3\,eV) of Ti are enhanced when using $p$-polarized 
photons for excitation~\cite{Ouardi2010}. The relative change in the intensity 
of peak E arising from $t_{1}$ states of Ni and Ti is larger in the superlattice 
sample (see difference curve in Fig.~\ref{fig:LDAD}(b)). This is due to the 
different overlying material in the two samples and therefore a increased 
contribution of states from Zr and Hf. In the bilayer sample the enhancement of 
the relative change in peak D at -3.06\,eV giving a sharper feature in the 
difference curve (Fig.~\ref{fig:LDAD}(a)) is caused mainly by changes of the 
cross sections for $t_{2}$ states of Ni similarly as it was observed previously 
for polycrystalline NiTiSn~\cite{Ouardi2011}. This is in a good agreement with 
the present case as the 30\,nm overlying layer mostly contributes to the overall 
signal obtained from the bilayer structure. From the polarization dependence it 
is also concluded that the {\it in-gap} states have $d$-type character.

%\section{Summary and Conclusions} %%%%%%%%%%%%%%%%%%%%%%%%%%%%%%%%%%%% In 
summary, the investigation of electronic properties of thin films as well as 
superlattices of NiTiSn and NiZr$_{0.5}$Hf$_{0.5}$Sn thermoelectric materials 
were performed by means of HAXPES. The polarization dependent HAXPES 
investigation allowed clearly to distinguish the states of different symmetry 
contributing to the total DOS in the valence band region in the pure NiTiSn and 
NiZr$_{0.5}$Hf$_{0.5}$Sn thin films. The impact of the different materials could 
even be resolved in the complex multilayered structures. Utilizing of NiTiSn as 
buffer layer for epitaxial growth of the different thin films and superlattices 
of both materials results in a high quality of the crystalline structure. The 
studies showed the appearance of {\it "in-gap"} $d$-states in both compounds 
that may be mediated by disorder at the interfaces and possible strain effects 
common for thin film structures. The {\it "in-gap"} states can serve as a tool 
for artificial tuning of the thermoelectric properties in thin films -- in 
particular an increase of the conductivity --, as was shown already for bulk 
materials.

%Acknowledgement %%%%%%%%%%%%%%%%%%%%%%%%%%%%%%%%%%%%%%%%%%%%%%%%%%%%%

\bigskip
%%%%%%%%%%%%%%%%%%%%%%%%%%%%%%%%%%%%%%%%%%%%%%%%%%%%%%%%%%%%%%%%%%%%%%%%%%%%%%%%
%\begin{acknowledgments}

Financial support by the DFG (Fe633/8-1 and Ja821/4-1 in SPP 1386) is gratefully 
acknowledged. HAXPES was performed at BL47XU of SPring-8 with approval of JASRI 
(Proposals No.~2011A1464, 2010A0017).

%\end{acknowledgments}

%

\end{document}